\documentclass[twocolumn,dvipsnames]{aastex62}

\usepackage{CJK}
\usepackage{framed}

\def\dim#1{{\rm\,#1}}
\def\Msun{{\rm\,M}_\odot}

\graphicspath{{./}{figures}}

\submitjournal{ApJ}

\shorttitle{}
\shortauthors{}

\begin{document}
\begin{CJK*}{UTF8}{gkai}

\title{Cosmic Reionization On Computers: Statistics, Physical Properties and Environment of Lyman Limit Systems at $z\sim6$}

\correspondingauthor{Jiawen Fan}
\email{johnfan@umich.edu}
\author[0009-0002-9477-886X]{Jiawen Fan（樊稼问）}
\affiliation{Department of Physics; University of Michigan, Ann Arbor, MI 48109, USA}

\author[0000-0003-0861-0922]{Hanjue Zhu (朱涵珏)}
\affiliation{Department of Astronomy \& Astrophysics; 
The University of Chicago; 
Chicago, IL 60637, USA}

\author[0000-0001-8868-0810]{Camille Avestruz}
\affiliation{Department of Physics; University of Michigan, Ann Arbor, MI 48109, USA}
\affiliation{Leinweber Center for Theoretical Physics; Department of Physics; University of Michigan, Ann Arbor, MI 48109, USA}

\author{Nickolay Y.\ Gnedin}
\affiliation{Theoretical Physics Division; 
Fermi National Accelerator Laboratory;
Batavia, IL 60510, USA}
\affiliation{Kavli Institute for Cosmological Physics;
The University of Chicago;
Chicago, IL 60637, USA}
\affiliation{Department of Astronomy \& Astrophysics; 
The University of Chicago; 
Chicago, IL 60637, USA}

\begin{abstract}
Lyman limit systems (LLSs) are dense hydrogen clouds with high enough HI column densities to absorb Lyman continuum photons emitted from distant quasars. Their high column densities imply an origin in dense environments; however, the statistics and distribution of LLSs at high redshifts still remain uncertain. In this paper, we use self-consistent radiative transfer cosmological simulations from the ``Cosmic Reionization On Computers'' (CROC) project to study the physical properties of LLSs at the tail end of cosmic reionization at $z\sim6$. We generate 3000 synthetic quasar sightlines to obtain a large number of LLS samples in the simulations. In addition, with the high physical fidelity and resolution of CROC, we are able to quantify the association between these LLS samples and nearby galaxies. Our results show that the  fraction LLSs spatially associated with nearby galaxies is increasing with the HI column density. Moreover, we find that LLSs that are not near any galaxy typically reside in filamentary structures connecting neighboring galaxies in the intergalactic medium (IGM). This quantification of the distribution and associations of LLSs to large scale structures informs our understanding of the IGM-galaxy connection during the Epoch of Reionization, and provides a theoretical basis for interpreting future observations.
\end{abstract}


\section{Introduction}\label{sec:intro}

As the first stars and galaxies form, photons emitted from these luminous sources travel through the Universe. These photons encounter hydrogen atoms residing in a variety of environments, from the low-density, diffuse intergalactic Medium (IGM) to denser filaments of cosmic large-scale structures to even denser gas in the interstellar Medium (ISM) of galaxies. 
The first ionizing photons rapidly ionize the neutral hydrogen in the diffuse IGM and turn it transparent to ionizing radiation (sometimes called Lyman Continuum, or LyC).  On the other hand, the denser regions remain progressively more neutral and more opaque to LyC. In observed absorption spectra of quasars and galaxies, the higher density and relatively more neutral regions manifest as Lyman Limit Systems (LLSs) \citep{Sargent89,Lanzetta91}.  LLSs are absorbers with the value of the hydrogen column density along the line-of-sight $N_{\rm HI}>1.6 \times 10^{17} \rm cm^{-2}$ (which corresponds to the unit optical depth at the Lyman limit) and less than about $2 \times 10^{20} \rm cm^{-2}$ \citep{Crighton19}.  The higher column density absorbers $N_{\rm HI}> 2 \times 10^{20} \rm cm^{-2}$ are traditionally called ``Damped Ly$\alpha$ Systems", or DLAs.  

Over the last decade, several groups have begun to explore how different physical processes and environmental factors affect the properties and distribution of LLSs.  For example, some studies examining the effects of radiation from stars on the distribution of LLSs at $z=3$ found minimal dependence on stellar feedback implementation \citep{yajima2012, altay2013}.  Other studies focused on understanding LLS properties in the context of their association with galaxies and streams at these lower redshifts \citep{fumagalli2011,rahmati2014,erkal2015}.  However, studies of LLSs at higher redshifts are sparse in the literature \citep{Songaila2010,Crighton19}, likely due to the previous limitations in available observational data for comparison. In particular, observational detection of LLSs at the highest redshifts requires more quasar spectra, particularly with complete spectral coverage of both the Ly$\alpha$ and the Lyman break \citep{Shukla16}. 

It is important to understand LLSs at high z. It is well established that LLSs contribute more to the ionizing photon mean free path (MFP) than the Ly$\alpha$ forest \citep[i.e.\ systems with $N_{\rm HI} \lesssim 1.6 \times 10^{17} \rm cm^{-2}$;][]{Miralda2003}, and LLSs are much more frequently seen than the denser and rarer DLAs. Therefore, LLSs are expected to play a particularly important role during cosmic reionization. As the ionized bubbles grow and merge, the embedded LLSs restrict the propagation of ionizing photons inside the ionized bubbles. LLSs thereby limit the MFP of the LyC photons, eventually causing local reionization along that line of sight to stop. An understanding of the nature of LLSs requires a combination of improved observations and simulation models.

The recent launch of JWST has opened a new window to study galaxy formation during the Epoch of Reionization. JWST near-infrared and mid-infrared bands enable observers to study the earliest galaxies and to probe galaxy formation to fainter magnitudes \citep{jades3,jades2,jades1,uncov1,aspire0,aspire1,glass0,glass1,ceers0,ceers1}. This new accessibility of galaxy formation data during the Epoch of Reionization is a major advancement compared with the historic paucity of high redshift LLS data.

Theoretical models, particularly cosmological simulations, are necessary complements to the deluge of observational data.  
Fortunately, the rapid progress in developing fast and accurate methods for modeling radiative transfer in cosmological simulations allows us to study the LLSs at high redshifts at an unprecedented level of detail.

Realistic models are important to our understanding of high-density absorbers \citep[e.g.][]{erkal2015}.  With galaxy formation physics, variations in the balance between heating and cooling processes can lead to a range of effects on the distribution of absorbers.  For example, more efficient cooling in gas clumps can lead to more compact dense gas distributions that background quasars are less likely to probe; these objects will fill the high-density tail of the true distribution of absorbers in the Universe but will suffer from an observational bias. On the other hand, energy feedback can expel high-density gas to larger radii, creating sufficiently dense gas for star formation in a ``positive feedback'' scenario \citep{Ishibashi2013,Luisi2021}. This kind of mechanism might potentially boost the number of intermediate-column-density absorbers at larger distances from galaxies. However, some studies on lower-redshift LLSs suggest LLS robustness to a realistic range of physical properties; star-forming gas in a galaxy will adjust such that outflows driven by feedback balance inflows due to accretion \citep{altay2013}.

The structure of this paper is as follows.  We describe the Cosmic Reionization on Computers (CROC) simulations and our methods to identify and quantify LLSs in Section~\ref{sec:methods}, present the statistics and properties of LLSs in Section~\ref{sec:results}, and summarize our results in Section~\ref{sec:summary}. 
\section{Methodology}\label{sec:methods}

\subsection{``Cosmic Reionization on Computers'': CROC Simulation}\label{sec:meth:CROC Simulation}
CROC simulations use the Adaptive Refinement Tree code \citep{Kravtsov99,Kravtsov02,Rudd08}. To account for the physics necessary to self-consistently model cosmic reionization, CROC simulations include gravity, gas dynamics, fully coupled (to gas dynamics) radiative transfer, atomic cooling and heating processes, molecular hydrogen formation, star formation, and stellar feedback. Full details of the simulation are described in the CROC method paper \citep{Gnedin14}.

For our analysis, we use three simulation runs with box length $L_{box}$ = 40 $h^{-1}$cMpc. The three runs are performed with exactly the same code and only differ in their random realizations of cosmological initial conditions. The simulations have a maximum spatial resolution of 100 pc in proper units. We generate 1000 lines of sight in each of the simulation boxes. These lines of sight start at random locations uniformly sampling the simulation volume, and they are oriented along random directions that sample the unit sphere uniformly. Each line of sight is sampled nonuniformly, fully tracing the underlying high resolution of the adaptively refined grid. We choose 100 $h^{-1}$cMpc for the length of the line of sight since it is a reasonable distance to use for the simulation boxes with $L_{box}$ = 40 $h^{-1}$cMpc, while providing a sufficient total length to allows us to find even rare absorbers.

\subsection{Identifying Lyman limit systems}\label{sec:meth:identification}

\begin{figure*}[htb!]
    \begin{minipage}{1\textwidth}
        \centering
        \includegraphics[width=.98\textwidth]{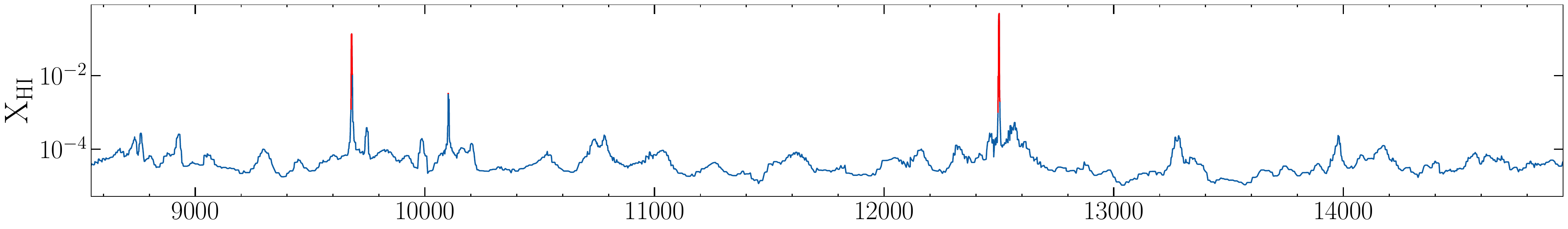}
    \end{minipage}\hfill
    \begin{minipage}{1\textwidth}
        \centering
        \includegraphics[width=1.0\textwidth]{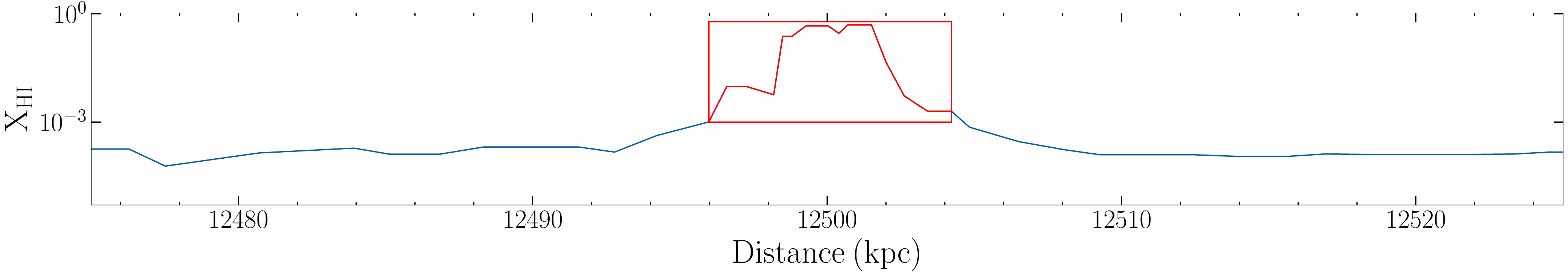}
\end{minipage}
\caption{Neutral fraction along an example line of sight.  The top panel shows two identified ``highly neutral'' regions in a single line-of-sight (red peaks), and the bottom panel zooms into the second peak with the red rectangle indicating the region of the line-of-sight above a threshold neutral fraction of $X_{\rm HI} = 0.001$ used to identify such regions.} \label{fig: example_LLS}
\end{figure*}

The observational signature of an LLS is a break at the Lyman limit (a drop in transmitted flux at 912\AA \ rest frame). In numerical simulations, we can simply look for regions with high enough neutral fraction ($X_{\rm HI}$). Figure~\ref{fig: example_LLS} shows the neutral fraction along one line-of-sight in our simulation.  Along each line of sight we search for regions with $X_{\rm HI}$ values larger than a given threshold, e.g.\ the region indicated by the red rectangle in the lower panel of Figure~\ref{fig: example_LLS}. We define the edges of these highly neutral regions to be at the intersection points between the threshold and the line of sight, e.g.\ the left and right sides of the red rectangle, which also corresponds to the ``piece'' of the LLS identified. The column density of the LLS is then simply the integral of the neutral hydrogen fraction along the line of sight between the two intersection points. However, we note that such a method for determining the {\rm HI} column density would not work reliably in the Ly$\alpha$ forest.

We use $X_{\rm HI} = 0.001$ as our fiducial threshold for highly neutral regions. This choice is somewhat arbitrary and needs to be tested.  Using a different threshold may change the column density of each LLS. In order to explore the artifacts induced by our choice of this threshold, we consider three different values for  $X_{\rm   HI}$ (0.0001, 0.001, and 0.003) and compare column density distributions for the three cases.

To calculate the column densities of LLSs, we count the LLSs in bins of natural log of the column densities, normalized by the total comoving path length sampled in the simulation ($\Delta l$) and the bin size. Historically, the observational data has been reported per unit ``absorption distance'' $\Delta X$. It is mostly a historical artifact and is defined as
\begin{equation}
dX = \frac{H_{0}}{H(z)}(1+z)^{2}dz,
\end{equation}
whereas we measure the comoving path length $\Delta l$ in the simulation. The conversion between $\Delta l$ and $\Delta X$ is straightforward:
\begin{equation}
dl/dX = \frac{(1+z)^{2}H_{0}}{c}.
\end{equation}

\begin{figure}[ht!]
\centering
        \includegraphics[width=0.99\columnwidth]{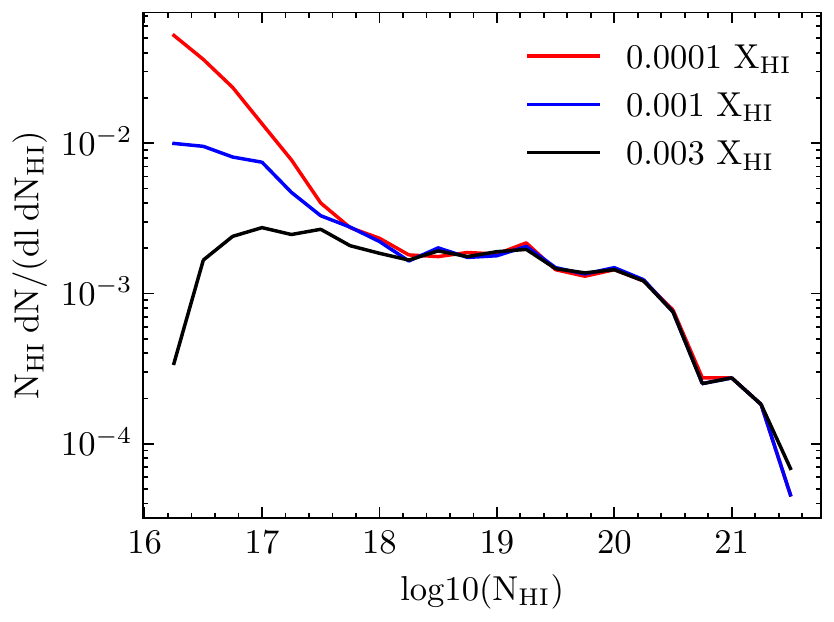}
\caption{Distribution of the number of LLSs at a given column densities, where we identify the LLSs at different threshold neutral fractions, $X_{\rm HI} = \{0.0001,0.001,0.003\}$. The distribution does not significantly vary with threshold choice for $N_{\rm HI}>3\times10^{17}{\rm cm}^{-2}$.} 
\label{fig: compare_xHI}
\end{figure}

We illustrate how the LLS column density distribution changes with varying the threshold $X_{\rm HI}$ in Figure~\ref{fig: compare_xHI}. The primary takeaway from the Figure is that the number of LLSs with $N_{\rm HI} > 3\times10^{17} \rm cm^{-2}$ is robust to the threshold choice.

\subsection{Galaxy Associations and Visualization of Lyman limit systems}\label{sec:meth:association}

A primary goal of this paper is to have a clearer understanding of where LLSs are located with respect to surrounding galaxies. We use the \textit{Rockstar} halo finder \citep{rockstar} to identify dark matter halos in our simulations. The well-defined center of a dark matter halo (i.e.\ the minimum of the gravitational potential) provides an accurate proxy for the location of each hosted galaxy, enabling spatial associations with LLSs. 

We define the ``association'' between LLSs and nearby galaxies using both the spatial locations and virial radii ($R_{vir}$) of the dark matter halos hosting these galaxies. We quantify the strength of associations using the quantity $f_{r<n\times R_{vir}} (n|n\in  \{1,2,3,4\})$, where $r$ is the distance between the location of the LLS neutral fraction peak and the center of the host dark matter halo of the nearby galaxy. For example, $f_{r<1\times R_{vir}} (1)$ corresponds to the fraction of LLSs' that sit inside $1$ $\times$ $R_{vir}$ of any galaxy. Higher $n$ values indicate weaker associations with any galaxy, since the LLS sits at larger distances with respect to the $R_{vir}$ of any galaxy.

Our definition for the LLS-galaxy association differs from some in the literature, e.g.\ \citet{rahmati2014}, who defined the associated galaxy as the closest detected galaxy to the LLS. \citep{Kohler07} showed that the latter approach is not as robust: for a given pair of a LLS and a nearby galaxy, one can frequently find a galaxy with a lower mass that is closer to the LLS, since lower mass galaxies are more numerous. And, as long as the numerical resolution of a simulation is sufficient to resolve such lower mass galaxies, these objects will be identified as the associated object to the LLS. As the result, the nearest neighbor identification is dependent on numerical resolution. In contrast, defining the LLS association with respect to the virial radius of the candidate galaxy automatically excludes all smaller galaxies within the virial radius of a larger one (i.e.\ satellite galaxies).

In addition, we note that the association of LLSs with satellite galaxies can be rather ambiguous.  First, satellites can be tidally truncated, so their virial radius is not always meaningful. Second, the satellites may have tidal tails that contain material from the satellite galaxy but are bound to the host; it is ambiguous whether the material is associated with a satellite or the host.  We choose to associate LLSs with the host galaxy through the virial radius association criterion in order to avoid this ambiguity. Additionally, we consider this approach more (although not completely) numerically robust and less dependent on numerical resolution. Granted, if a given LLS is not associated with any galaxy for a given value of $n$, this does not imply that there is no association for a lower mass galaxy that is not resolved in the simulation.

To help visualize the LLSs, we generate gas density projection plots around some LLSs in simulation box snapshots using the ``yt'' simulation analysis package \citep{Turk11}. In these projection plots, one can visually identify structures surrounding the LLS including galaxies and connective filaments.

\section{Results}\label{sec:results}

\subsection{Statistics of LLSs}
\label{sec:results:lls_stats}

\begin{figure*}
    \centering
    \begin{minipage}{\columnwidth}
        \centering
        \includegraphics[width=0.99\columnwidth]{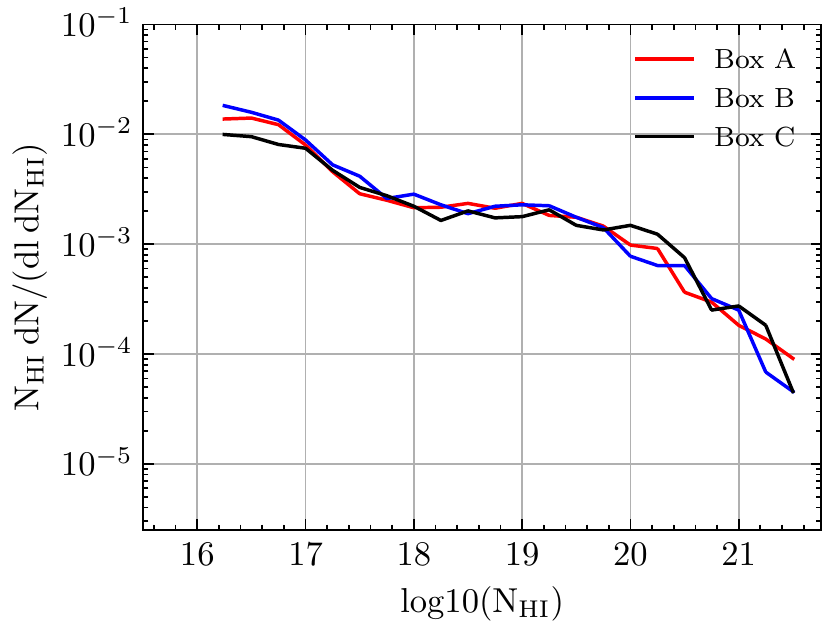} 
    \end{minipage}
    \begin{minipage}{\columnwidth}
        \centering
        \includegraphics[width=0.99\columnwidth]{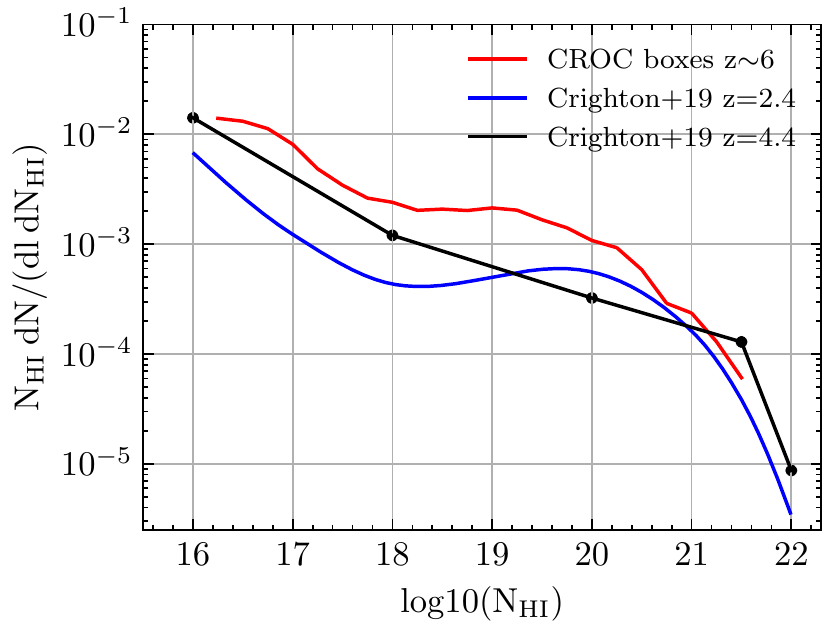}
        
\end{minipage}
\caption{Left: column density distribution of LLSs for three different 40 $h^{-1}$cMpc boxes. Right: comparison between the observational fit at $z=2.4$, $z=4.4$ \citep{Crighton19} and our simulations.}
\label{fig:column_density_distribution}
\end{figure*}

We use the column density distribution of LLSs defined in Section~\ref{sec:meth:identification} to compare column density distribution of LLSs for CROC simulations. In Figure~\ref{fig:column_density_distribution}, the left panel shows the individual distributions from each of the three different realizations of the CROC simulations as an estimate of the sample variance in a finite simulation volume. Here, we see that the column density distributions of LLSs in CROC exhibit an overall trend where the LLS number density decreases with increasing column density. Overall, the three simulation boxes produce similar column density distributions within the level of the numerical noise with which we are able to measure them.

The right hand panel shows the column density distribution of LLSs from the combined three CROC boxes, alongside the fits to observational data from \citet{Crighton19} at $z=2.4$ and $z=4.4$. We see that our simulation result and the observational fits have similar shapes, with CROC simulations having more LLSs with $16 < \log_{10} (N_{\rm HI}) < 19 $. The differences between the column density distributions of \citet{Crighton19} data and the average distribution across all CROC boxes are reasonable given the limited observational data and modeling uncertainties.  First, we expect potential variations up to an order of magnitude for the distribution at the lowest column density bins ($\log_{10}N_{HI} < 17.5$) that depend on the neutral fraction threshold criterion for identifying LLSs in our simulations (see Figure~\ref{fig: compare_xHI}).  Next, we see that the observational data in the highest column density bins ($\log_{10}N_{HI}\gtrsim 20$) do not exhibit any strong evolutionary trend.  The measurements from CROC are consistent with this behavior.  The overall larger amplitude in column density distribution of LLSs in CROC at $z\sim6$ is consistent with the suggested evolution from observations indicating an increasing normalization with redshift.  Some increase in normalization with redshift is plausible since we expect more dense absorbers to exist during reionization, with a larger fraction of neutral hydrogen present compared with lower redshift. It is sufficient for our purpose that our simulation measurements capture the dominant trends in the data.

\subsection{Association of LLSs with Galaxies}

\begin{figure}[hb]
\centering
        \includegraphics[width=0.90\columnwidth]{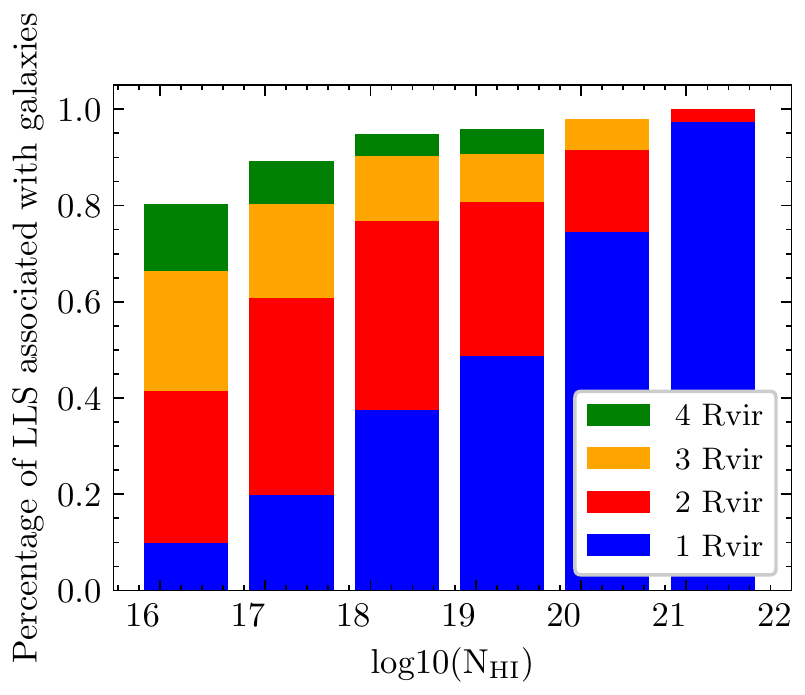}
\caption{The fraction of Lyman limit systems within $n\times R_{vir}$ of any galaxies, with $n\in{1,2,3,4}$. Blue, red, orange, and green indicate respective fractions of Lyman limit systems within that distance of any galaxy. We see that higher column density systems have progressively stronger associations with galaxies.}
\label{fig:fvir}
\end{figure}

A primary goal of this paper is to provide theoretical predictions of how LLSs at $z\sim6$, near the end of cosmic reionization, are associated with galaxies and nearby structures.  
In order to quantify the association of Lyman limit systems and galaxies in our simulation, we show in Figure~\ref{fig:fvir} the fraction of LLSs whose neutral fraction peaks are within $n\times R_{vir}$ of any galaxy, i.e. the $f_{r<n\times R_{vir}} (n| n\in\{1,2,3,4\})$ quantity we discussed in Section~\ref{sec:meth:association}. The fraction is for all LLSs from each of our 40 $h^{-1}$cMpc simulation boxes. The overall trend with the neutral hydrogen column density is very clear: the damped Ly$\alpha$ systems of $\rm \log_{10}\it N_{HI}>$ 20 are predominantly located inside galactic halos. At lower column densities the association of LLSs with galaxies weakens, with a larger fraction of these LLSs located further than $2\times R_{vir}$ from any galaxy. As we go down to the Ly$\alpha$ forest range, $N_{\rm  HI}<10^{17}{\rm cm}^{-2}$, only 10\% of these systems are within $1\times R_{vir}$ of a galaxy center and only 40\% are within $2\times R_{vir}$.

\begin{figure*}[htb!]
    \begin{minipage}{0.49\textwidth}
        \centering
        \includegraphics[width=0.99\textwidth]{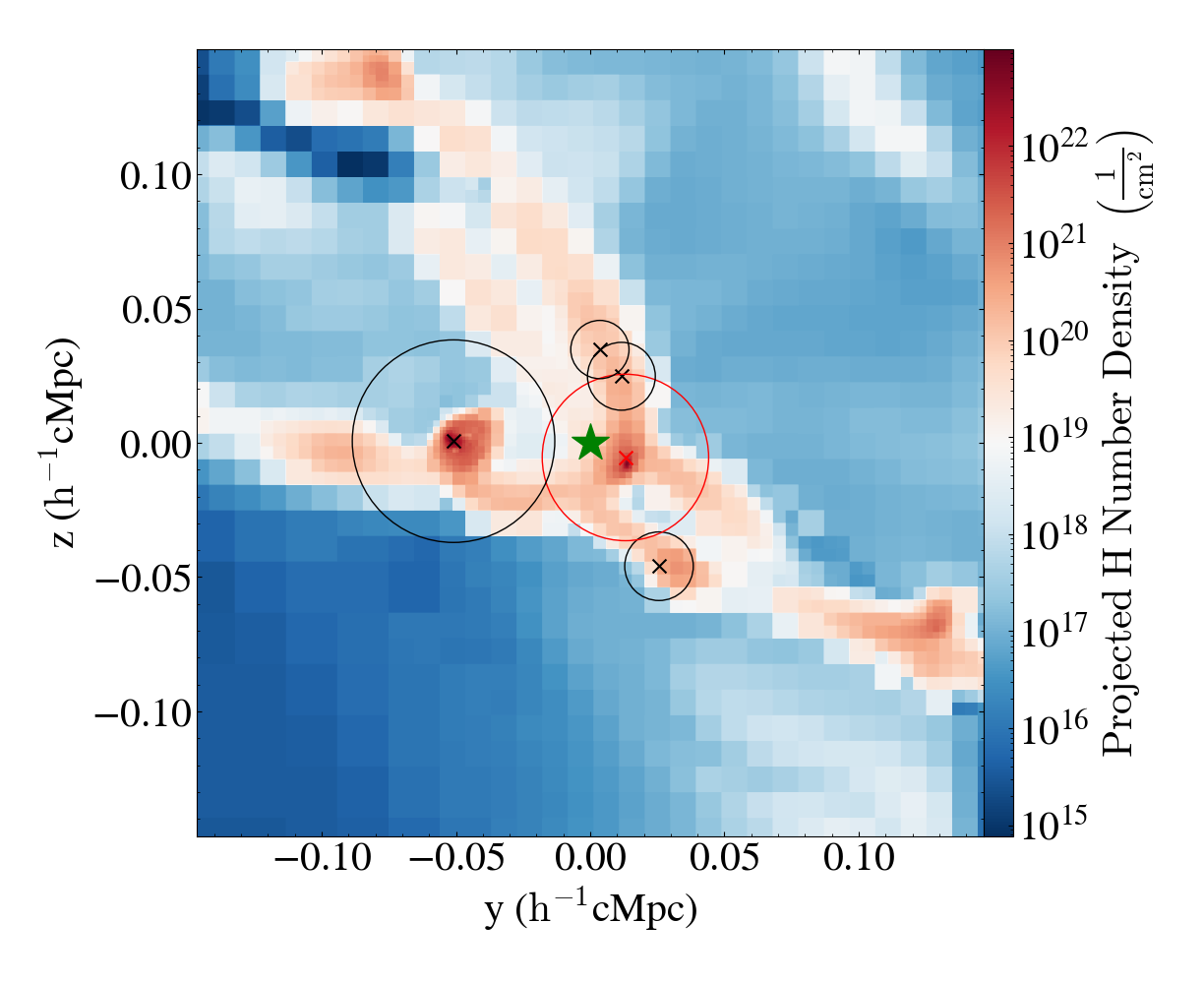}
    \end{minipage}\hfill
    \begin{minipage}{0.49\textwidth}
        \centering
        \includegraphics[width=0.99\textwidth]{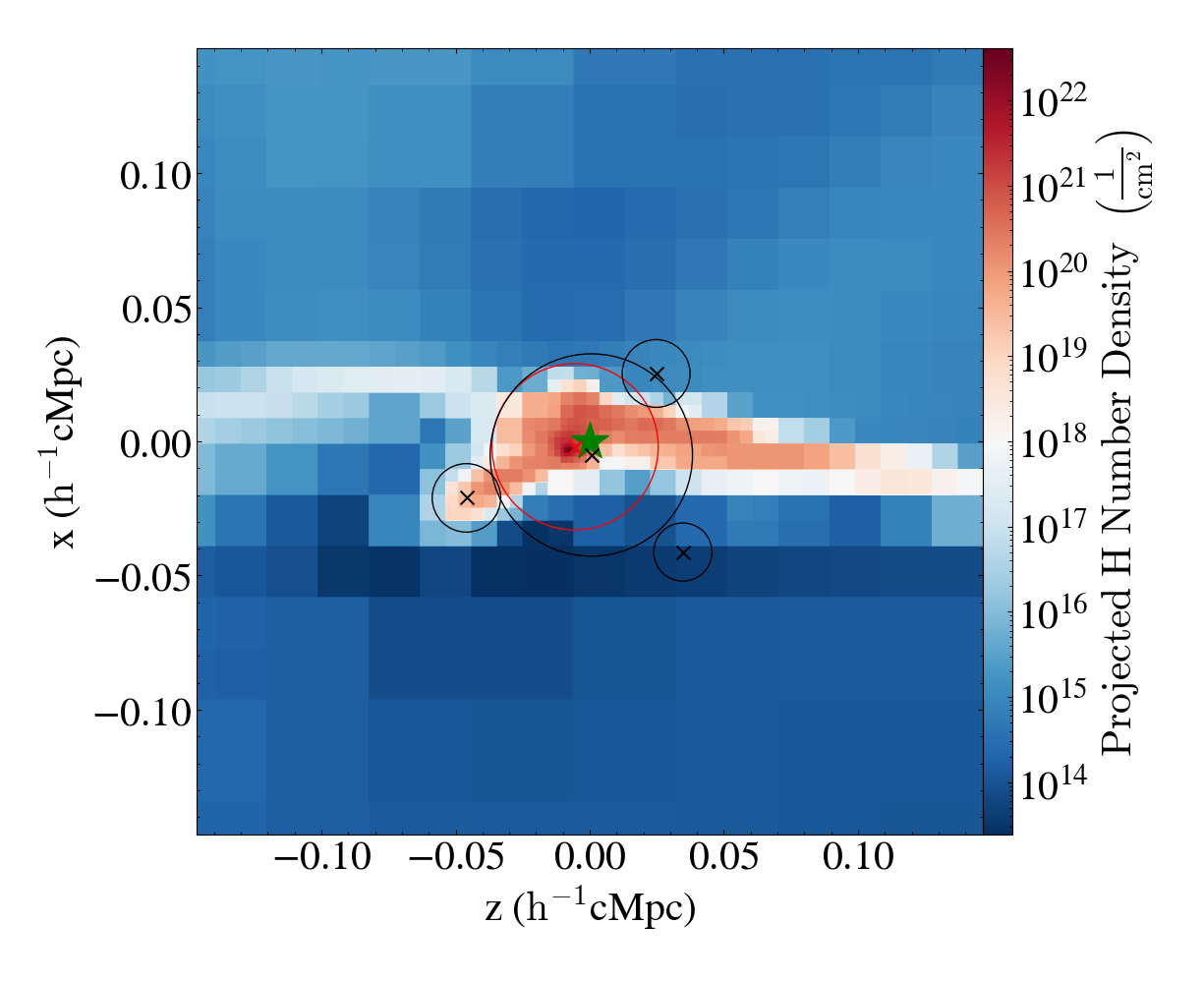}
\end{minipage}
\caption{Projected neutral hydrogen maps centered around a Lyman limit system with $N_{  HI}=10^{19.4}{\rm cm}^{-2}$. The projection depth of the graph is $\sim$ 10 kpc, corresponding to the characteristic size of the Lyman limit system.  Left: zy-projection plane.  Right: xz-projection plane. The green star indicates the position of the Lyman limit system; the black circles center on galaxies and indicate their corresponding dark matter halo virial radii; the red circle corresponds to the galaxy that contains the Lyman limit system within its virial radius. Most of the high column density Lyman limit systems reside within one virial radius of a galaxy.}\label{fig:LLS_within_Rvir}
\end{figure*}

Of the population of highest column density LLSs (i.e.\ those with $10^{19}{\rm cm}^{-2}<N_{\rm  HI}<10^{20}{\rm cm}^{-2}$), 80\% are within $2\times R_{vir}$ of some galaxy. Figure~\ref{fig:LLS_within_Rvir} shows the projected neutral hydrogen density around a LLS with the column density of $N_{HI}=10^{19.4}{\rm cm}^{-2}$, located within $1\times R_{vir}$ of the virial radius of a galaxy. The left panel shows the projection in the y-z plane, and the right in the x-z plane. The depth of projection corresponds to the estimated size of the LLS ($\sim$10~kpc).

We indicate the LLS location with a green star, and surrounding dark matter halo centers with their virial radii respectively in black crosses and black circles. The galaxy associated with the LLS has a dark matter halo mass and virial radius of $M=2.8 \times 10^9 \Msun$ and $R_{vir} = 6.8 \rm\ kpc$ respectively; we show it in red.  

\begin{figure*}[htb]
    \begin{minipage}{0.49\textwidth}
        \centering
        \includegraphics[width=0.99\textwidth]{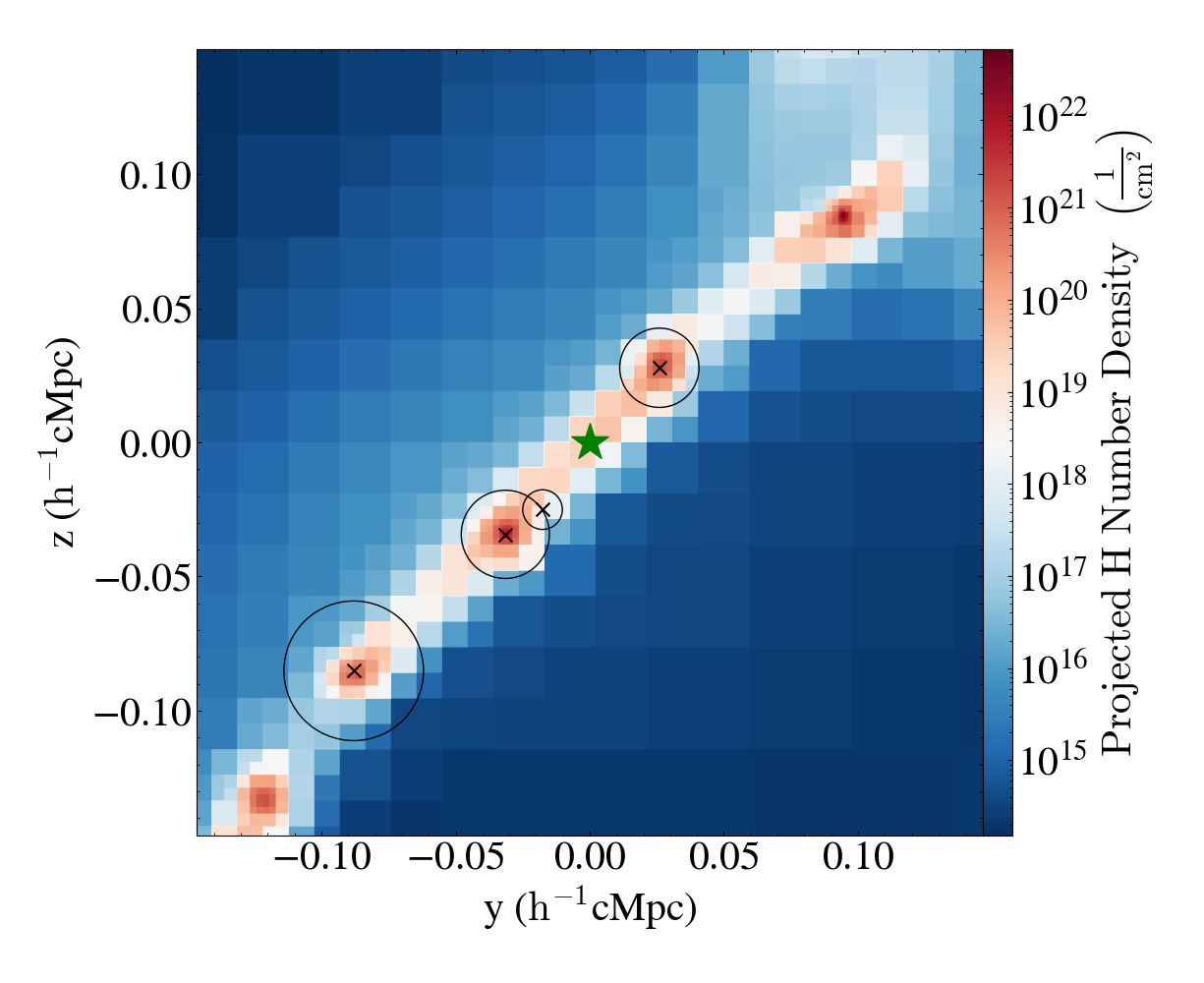}
    \end{minipage}\hfill
    \begin{minipage}{0.49\textwidth}
        \centering
        \includegraphics[width=0.99\textwidth]{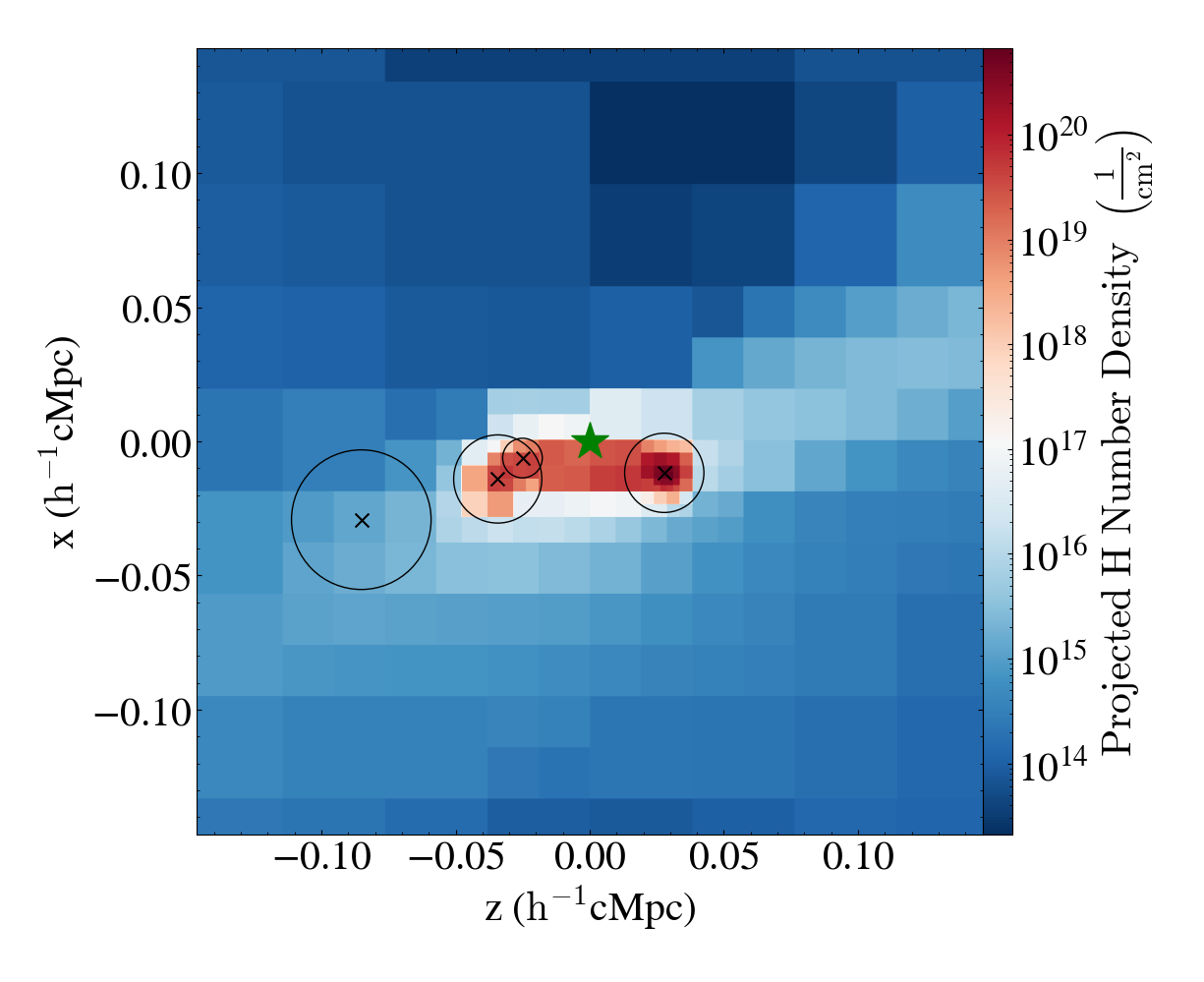}
\end{minipage}
\caption{Projected neutral hydrogen maps centered around an example Lyman limit system weakly associated with any galaxy. The projection depth is $\sim$ 10 kpc. Markers indicate the same properties as in Figure~\ref{fig:LLS_within_Rvir}.  This Lyman limit system is well outside of 2 times the virial radius of any galaxy.  For illustration purposes, we annotate galaxies up to 5 virial radii away from the Lyman limit system.}
\label{fig:LLS_4_Rvir}
\end{figure*}

The remaining 20\% of the highest column density LLSs are outside of $2\times R_{vir}$ of any galaxy.  While these are not closely associated with any galaxy, we find that they are associated with overdense filamentary structures in the large-scale surrounding environments. Figure~\ref{fig:LLS_4_Rvir} shows the projected neutral hydrogen map surrounding such a system, with similar projections and color mapping as Figure~\ref{fig:LLS_within_Rvir}.  Here, we see that the LLS sits in a filamentary bridge connecting galaxies, but lies well outside twice the virial radius of nearby galaxies. 

\subsection{Physical Properties of LLSs}

\begin{figure*}[htb]
    \centering
        \includegraphics[width=.95\textwidth]{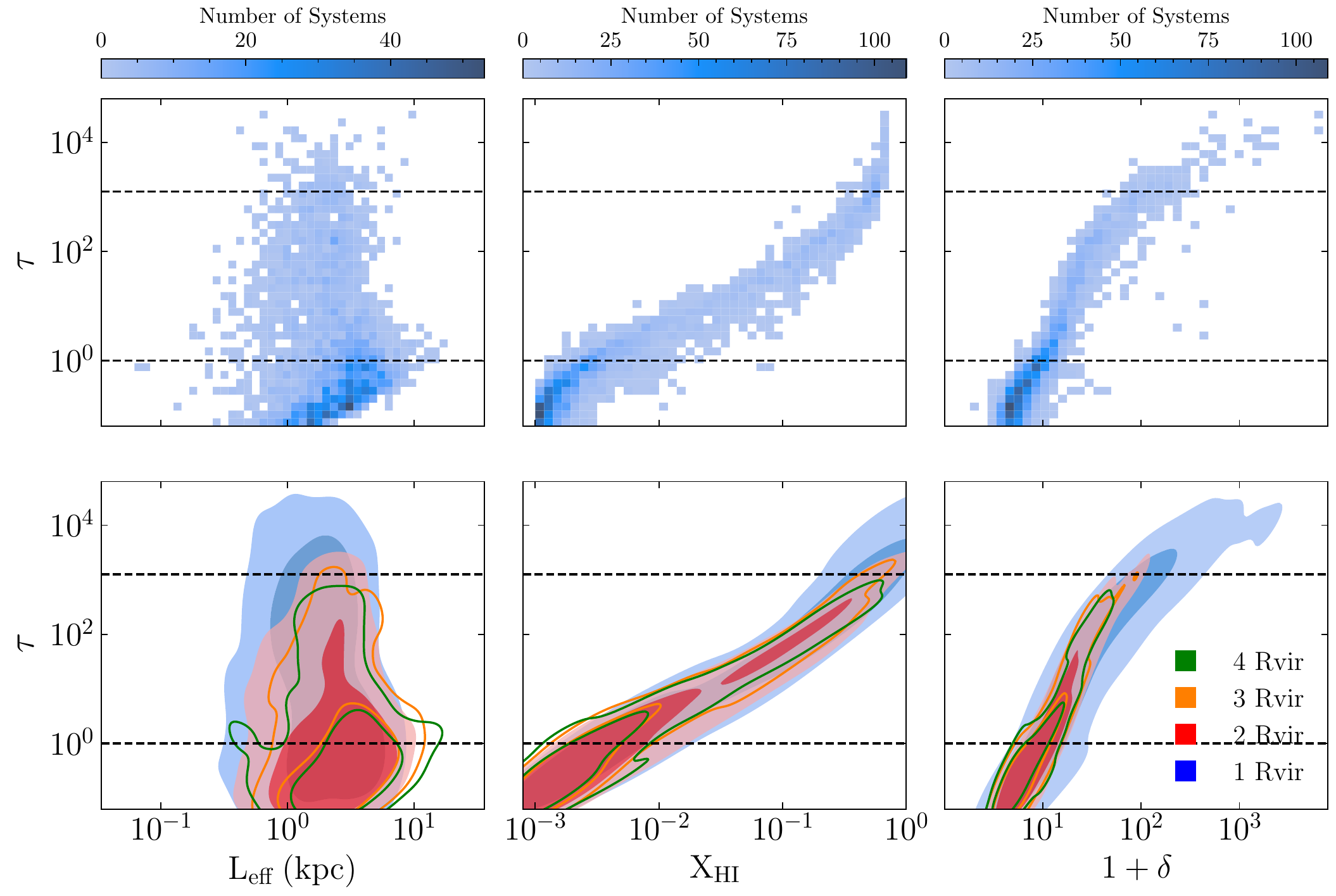}
\caption{Lyman limit system optical depth vs various physical properties of the Lyman limit systems in CROC. From left to right: characteristic size, neutral fraction, and gas overdensity, respectively. The top panels are two-d histogram plots that visualize the number density distribution of data points for each LLS. The color bar above indicates the number of data points in each square bin. The bottom panels show the same distribution of systems in subsets of association strength with any galaxy.  The blue, red, orange, and green contours respectively correspond to systems within 1, 2, 3, and 4 times the virial radius of any galaxy. For each color, the contours correspond to 1-$\sigma$ and 2-$\sigma$ regions of the distribution.}

\label{fig:physical_quantities}
\end{figure*}

In this section, we investigate the relationship between the optical depth of LLSs and their various physical properties in CROC simulations. The LLS physical properties of interest here are: optical depth, characteristic size, neutral fraction, and density. We measure all physical properties for each LLS at the peak neutral fraction location of the LLS along the line of sight.

Optical depth describes opacity of the LLSs and is directly proportional to the column density. Optical depth is defined as:
\begin{equation}
\tau = \sigma_{\rm LL} N_{\rm   HI},
\end{equation}
where $\sigma_{\mathrm{LL}} = 6.3 \times 10^{-18} \ \rm cm^2$ is the hydrogen photoionization cross section at the Lyman limit. 

We define the characteristic size of a LLS as:
\begin{equation}
L_{\rm eff} = N_{\rm HI}/n_{\rm HI},
\end{equation}
where $n_{\rm HI}$ is the HI number density at the peak neutral fraction location of the LLS along the line of sight. If the HI number density at this peak had a Gaussian shape, this definition would closely correspond to the FWHM.  

The top panel of Figure~\ref{fig:physical_quantities} illustrates the two-dimensional distributions of these four quantities for the simulated LLSs. The color bar on top of each plot represents the number of data points inside each bin. The horizontal dashed lines show the $\tau$ range corresponding to the LLS column density range, although we note that the numbers are arbitrarily defined. 

We see from the middle and right panels that $\tau$ increases with the peak neutral fraction and the peak gas density of the LLSs. This means that higher column density LLSs are more neutral and denser at the center. However, there is not much of a trend between $\tau$ and $L_{\rm eff}$ - i.e., at a fixed column density (or, equivalently, $\tau$), some LLSs are compact dense gas clumps while others are more spatially extended but less dense.  Most LLSs, however, have sizes between 1 and $\sim$5 physical kpc. Note, that for LLSs with $\tau \ge$ 100, their neutral fractions are close to unity. We also know that these high column density LLSs are strongly associated with nearby galaxies from Figure~\ref{fig:fvir}. This means that despite their proximity to ionizing sources, high column density LLSs still have highly neutral centers that are not affected by the ionizing photons.

It is also instructive to separate LLSs based on the strength of their association to galaxies, and plot the physical quantities in the bottom panel of Figure~\ref{fig:physical_quantities}. The colors we use here are the same as in Figure~\ref{fig:fvir}.  Blue, red, orange, and green correspond to systems within 1, 2, 3, and 4 times the virial radius of any galaxy. The contour regions correspond to the 1-$\sigma$ and 2-$\sigma$ regions of the distribution for that subset of the LLSs.   We chose to fill in the blue and red contours to visually clarify the distribution at each association strength.  The strength of association with galaxies does not appear to have any significant effect on the physical properties of LLSs at $z\sim6$.

\subsection{Shapes and Orientation of Lyman limit systems}
\label{sec:llsdistribution}

\begin{figure}
\centering
        \includegraphics[width=0.9\columnwidth]{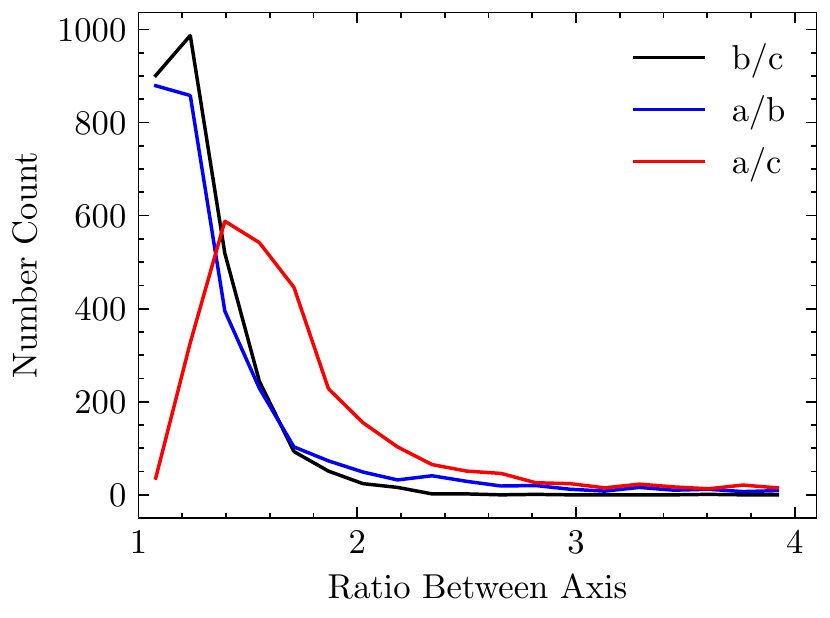}
\caption{Distribution of shape axis ratios measured from our Lyman limit system samples.  Both $b/c$ and $a/c$ peak near 1.3, indicating that most of the systems have gas distributions that deviate from spherical.  The ratio of the semi-major axis to the semi-minor axis, $a/c$ exhibits a longer tail and skews large, corresponding to the particularly elongated systems.}
\label{fig:ratio_graph}
\end{figure}

LLSs at a fixed column density can come from both (a) systems of lower average column density seen along a particular direction with higher gas column density and (b) systems of higher average column density seen along a particular direction with low gas column density.  Such interdependencies imply that the LLS column density distribution quantified in Figure~\ref{fig:column_density_distribution} reflects both the distribution of LLSs orientations in space and the distribution of their shapes.  This motivates us to study the shapes and orientations of LLSs.

\begin{figure}[bt!]
\centering
        \includegraphics[width=0.9\columnwidth]{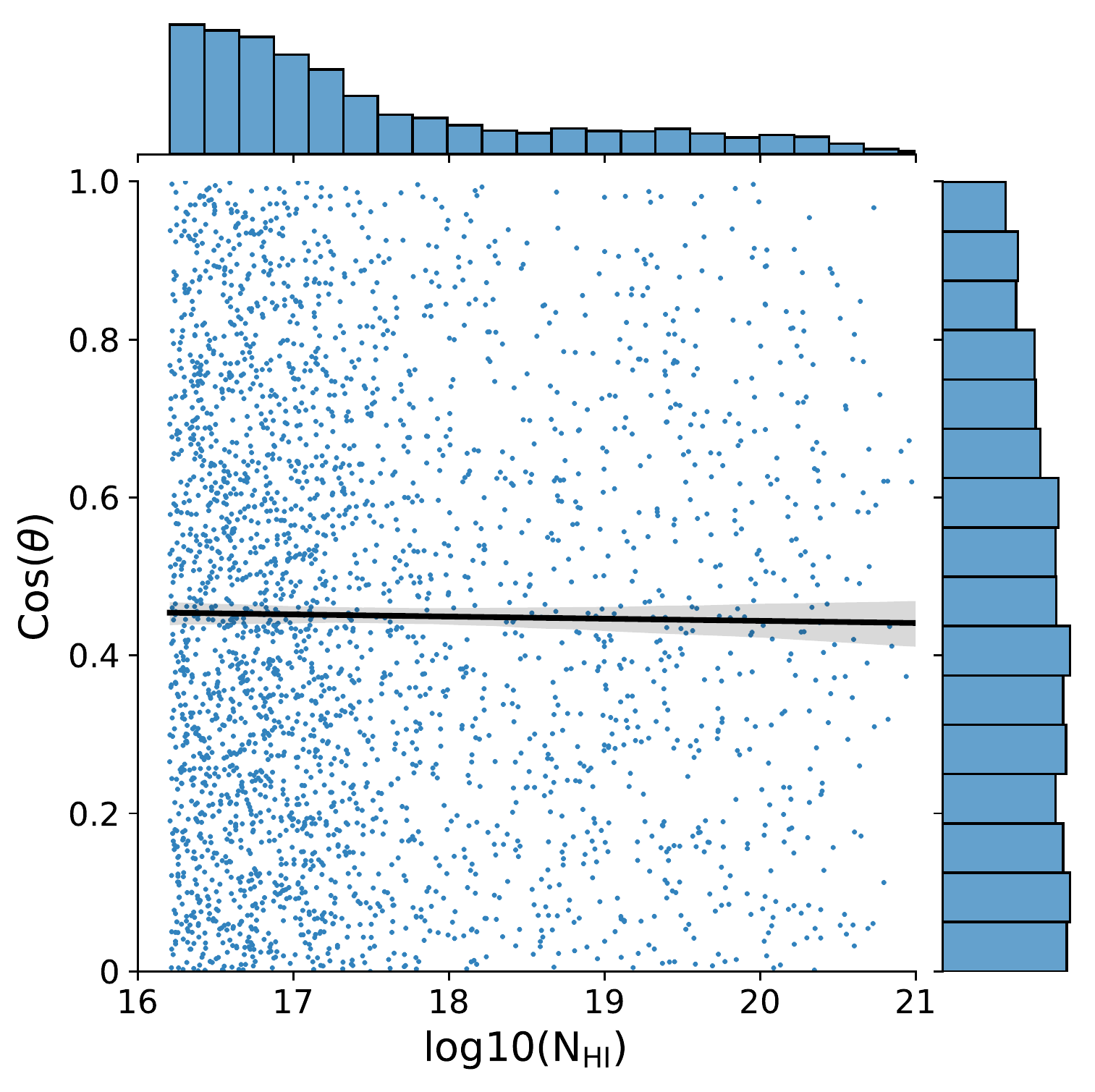}
\caption{Blue points: Alignment between the line of sight and the principal axis of each LLS, $\cos\theta$, as a function of the measured column density for that system, $\log_{10}N_{HI}$.  Black line:  Regression line with shaded regions corresponding to the confidence interval.  Blue histograms on axes: Corresponding marginal histograms for each quantity.  There is no apparent trend in alignment with the column density.}
\label{fig:cos_theta}
\end{figure}

In order to explore the distribution of shapes of LLSs and any potential bias associated with their non-sphericity, we calculate the moment of inertia tensors of all simulated LLSs within 5 proper kpc from the location of the density peak in the LOS.  We then compute the axis ratios of equivalent ellipsoids (i.e.\ an ellipsoid with the same moment of inertia tensor). We choose 5~kpc based on the left panel of Figure~\ref{fig:physical_quantities}, which shows that $L_{\rm eff}\lesssim 5\,{\rm pkpc}$ for the vast majority of all LLSs. We show the resulting distribution of axis ratios of equivalent ellipsoids for our simulated LLS sample in Figure~\ref{fig:ratio_graph}. 

\begin{figure}[bt!]
\centering
        \includegraphics[width=0.9\columnwidth]{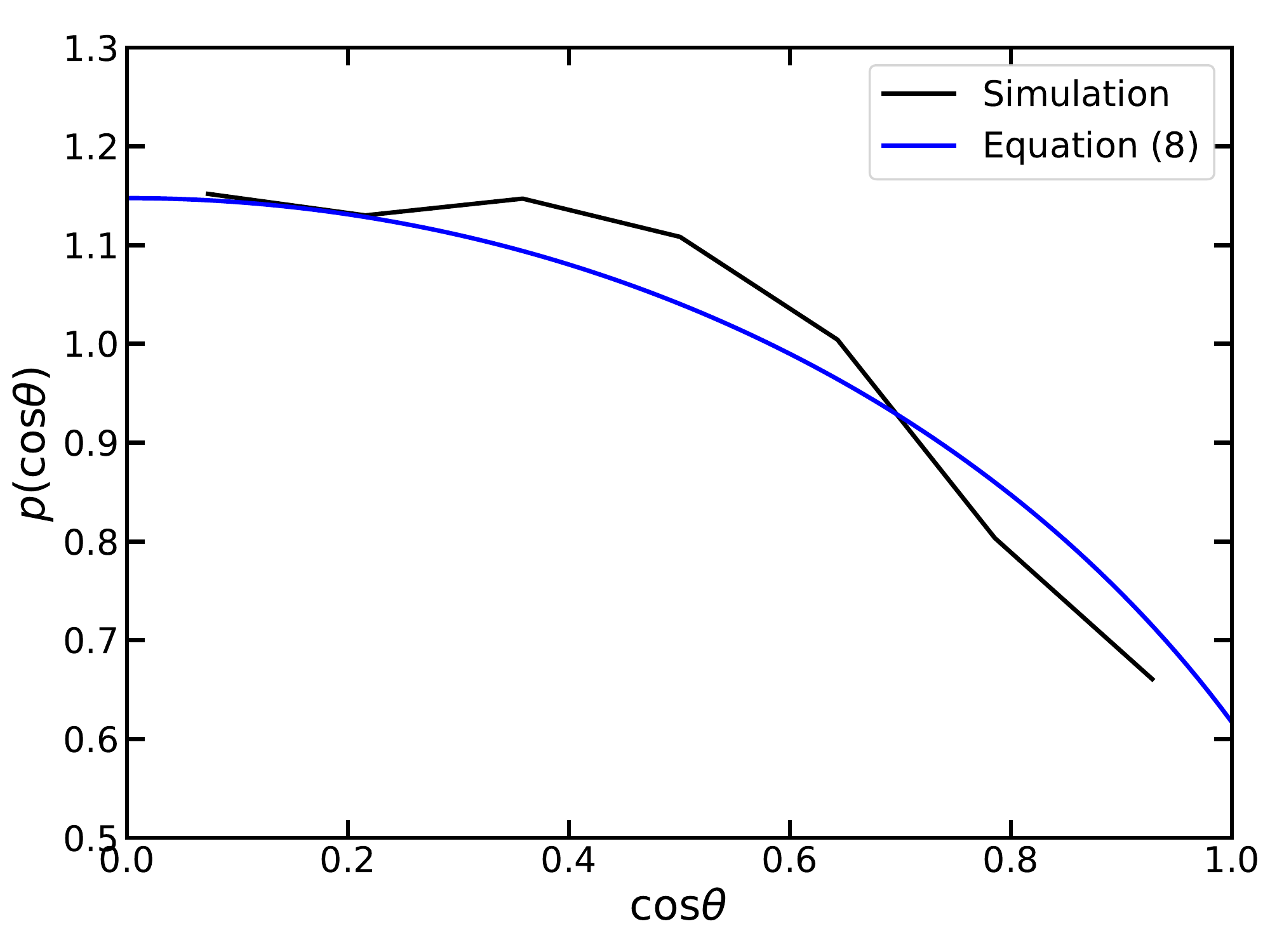}
\caption{Probability density function of $\cos(\theta)$ for our simulated sample of LLSs ($N_{\rm HI}>10^{16}\dim{cm}^{-2}$, the black line - note that it is this cut that makes this distribution $\mu$-dependent). The blue line shows a simple model from Equation (\ref{eq:fnm}). This figure serves as a more quantitative version of the y-axis histogram from Fig. \ref{fig:cos_theta}. }
\label{fig:eigen_angle}
\end{figure}

Since LLSs are not completely spherical, lines of sight through an LLS may only capture a threshold density peak with preferred orientations to the LLS principle axes. In particular, if a line of sight aligns with the longest principal axis of the LLS (which corresponds to the smallest eigenvalue of the moment of inertia tensor), the observed column density will bias high compared to the average for any possible line of sight.  Specifically, $\langle N_{\rm   HI}\rangle \propto X_{\rm   HI}(1+\delta)L_{\rm eff}$ defined by the physical quantities shown in Figure~\ref{fig:physical_quantities}. 
In Figure \ref{fig:cos_theta}, we show the distribution of (cosines of) angles, $\cos\theta$, between the longest principal axis of the moments of inertia tensor and the LOS direction for all LLSs and plot this against the column density. There is a mild trend of more LLSs with smaller angles, resulting in the mean $\cos\theta$ being slightly lower than 0.5. Figure \ref{fig:eigen_angle} shows the distribution of $\cos\theta$ for all LLSs with \draft{$N_{\rm   HI}>10^{16}\dim{cm}^{-2}$}  (a more quantitative representation of the y-axis histogram from Figure \ref{fig:cos_theta}). This is, however, what is expected for approximately ellipsoidal shapes. Namely, for a homogeneous ellipsoid with axes $a$ and $c$ (here, for simplicity, we adopt $b=c$) a column density along an angle $\theta$ with respect to the major axis is $N = n\times l$, where $n$ is the density and $l$ is determined from the equation,
\begin{eqnarray}
    \frac{l^2\cos^2\theta}{a^2} + \frac{l^2\sin^2\theta}{c^2} = 1,    
\end{eqnarray}
or 
\begin{eqnarray}
   l = \frac{ac}{\sqrt{a^2+(c^2-a^2)\cos^2\theta}}. 
\end{eqnarray}

If the distribution of LLSs over the column densities and angles $\theta$ (shown in Figure~\ref{fig:cos_theta}) is $f(N,\mu)$ (where we use a short-hand notation $\mu\equiv\cos\theta$), then,
\begin{equation}
    f(N,\mu) = \int d N_a f(N_a) \delta\left(N - \tilde{N}(N_a,\mu)\right),
    \label{eq:fnm}
\end{equation}
where $N_a \equiv n\times a$ is the column density along the major axis, $f(N_a)$ is the column density distribution with respect to $N_a$, and $\tilde{N}(N_a,\mu)$ is the function that gives a value of $N$ for a given values of $N_a$ and $\mu$. Here we also assume that $f(N_a)$ is independent of $\mu$, since $f(N_a)$ is intrinsic to the system and, hence, independent of the location of the observer, i.e.\ $f(N_a,\mu) = C f(N_a)$, with $C = 1$ since $\mu$ goes from 0 to 1. This derivation serves as a mere illustration, so here we also assume that the ratio $c/a$ is constant, so that there is no additional distribution over $c$ for a fixed $a$.

From Equation (\ref{eq:fnm}) it immediately follows that
\begin{eqnarray}
    f(N,\mu) & = &\int d N_a f(N_a) \frac{dN_a}{dN}\delta\left(N_a - \tilde{N}_a(N,\mu)\right) \nonumber\\
             & = &f(N_a) \frac{\partial\tilde{N}_a}{\partial N}\nonumber\\
             &= &f(N_a) \frac{\sqrt{1+(\alpha^2-1)\mu^2}}{\alpha},
\end{eqnarray}

with $\alpha\equiv c/a$. Here $\tilde{N}_a(N,\mu)$ is the function that gives a value of $N_a$ for a given values of $N$ and $\mu$, i.e.\ the inverse function to $\tilde{N}(N_a,\mu)$. We consider this model distribution over $\mu$ for $N_{\rm  HI}>10^{16}\dim{cm}^{-2}$\footnote{Note that the threshold in $N_{\rm HI}$ makes the distribution $\mu$-dependent. If, instead, we chose the sample as all column densities above some $N_a$, the blue line in Fig.\ \ref{fig:eigen_angle} would be flat.} and the value of $\alpha=0.54$, the average value from Figure~\ref{fig:ratio_graph}, and overplot it in Figure~\ref{fig:eigen_angle} (blue dashed line) to compare against the distribution in alignment from our simulated sample (black solid line). The agreement between the simulation results and the simple model suggests that there is no significant orientation-dependent bias present.  The small deviations between the two lines cannot be considered significant given the simplicity of the analytical model.

\section{Summary and Discussion}\label{sec:summary}

We use the state-of-the-art reionization simulation, CROC, to investigate the statistics, environment and physical properties of Lyman limit systems at $z\sim6$. We cast a total of 3000 lines-of-sight to simulate synthetic quasar sight lines. Along these sight lines, we are able to identify LLSs in the simulated IGM.

Our main results are:
\begin{enumerate}
    \item Galaxy distribution largely dictates where we will find most of the LLSs. The highest column density self-shielded absorbers, such as damped Ly$\alpha$ systems, have the strongest association with galaxies. These are typically located within one virial radius of a galaxy. 
    \item Lower column density LLSs may be located further from the nearest galaxy, but even the systems with column densities as low as $10^{16}\,{\rm cm}^{-2}$ are mostly (80\%) located within four virial radii from the nearest galaxy, with the majority (50\%) of LLSs being located within just two virial radii from the nearest galaxy.
    \item Rare LLSs that are not located close to any galaxy are preferentially found along the filament connecting more distant galaxies.
    \item The column density of LLSs strongly correlates with the gas density and the hydrogen neutral fraction. Their characteristic sizes, however, show little correlation.
    \item The gas distributions around LLSs tend toward a slightly elongated gas distribution, with the peak distribution in axis ratio being $a/c\sim1.4$.  Despite this slight preference, our mock observed selection of LLSs remain unbiased: they are consistent with 
    every LLS being sampled fairly along all possible directions. 
\end{enumerate}

One of the physical properties of the LLSs that we measure is the effective size.  The effective size of the LLSs provides a reference point. The effective size of the LLSs is comparable to the sizes of galactic disks and smaller than virial radii of galactic halos at this redshift. Therefore, it is likely that LLSs are not largely ISM structures as DLAs are, which would make the effective size comparable to the disk scale height.  But, LLSs could be tidal tails from disrupted satellites or intergalactic filaments, as Figures~\ref{fig:LLS_within_Rvir} and \ref{fig:LLS_4_Rvir} seem to indicate.

In our fourth point above, we note that the strong correlation with density is the result of the effective size being weakly correlated with the column density. In general, column densities and physical densities do not need to be correlated. For example, if all LLSs were coming from pancakes or filaments, there would be no or weak $n^{1/3}$ correlation with density. 

We emphasize that the last point is nontrivial.  For a random line of sight to exhibit a density peak above our selection threshold, we would expect some dependence on the exact shape of the column density distribution.  In our mock observed distribution, it is just as likely for the more numerous lower column density systems to be seen along their major axis directions as for rarer higher column density systems to be seen along a direction more aligned with their minor axes.

\acknowledgments

The authors thank the anonymous referee for suggestions and comments that improved this manuscript. J.F. was supported by the Summer Undergraduate Research Experience fund from the University of Michigan's Physics Department. C.A. acknowledges support from the Leinweber Center for Theoretical Physics.  This work was supported in part by the NASA Theoretical and Computational Astrophysics Network (TCAN) grant 80NSSC21K0271.
This manuscript has been coauthored by Fermi Research Alliance, LLC under Contract No. DE-AC02-07CH11359 with the U.S. Department of Energy, Office of Science, Office of High Energy Physics. This work used resources of the Argonne Leadership Computing Facility, which is a DOE Office of Science User Facility supported under Contract DE-AC02-06CH11357. An award of computer time was provided by the Innovative and Novel Computational Impact on Theory and Experiment (INCITE) program. This research is also part of the Blue Waters Sustained Petascale computing project, which is supported by the National Science Foundation (awards OCI-0725070 and ACI-1238993) and the state of Illinois. Blue Waters is a joint effort of the University of Illinois at Urbana-Champaign and its National Center for Supercomputing Applications. This work was completed in part with resources provided by the University of Chicago Research Computing Center. This research was supported in part by the National Science Foundation under Grant No.\ NSF PHY-1748958. N.G. and H.Z. thank KITP at the University of California, Santa Barbara, for hospitality during a portion of this work.

\bibliographystyle{apj}
\bibliography{main}

\end{CJK*}

\end{document}